\documentclass[12pt]{article}
\usepackage{amsfonts}
\usepackage{amsmath}
\usepackage{amssymb}
\usepackage{axodraw}
\usepackage{mathptmx}
\usepackage[dvips]{color}
\usepackage{epsfig}
\usepackage{graphicx}
\newcommand{\calA}{{\cal A}}

\newcommand{\calL}{{\cal L}}

\newcommand{\calU}{{\cal U}}

\newcommand{\Etilde}{\tilde{E}}
\newcommand{\Ftilde}{\tilde{F}}
\newcommand{\Utilde}{\tilde{\calU}}

\newcommand{\tr}{\textrm{tr}}

\begin{document}
\baselineskip=16pt

\pagenumbering{arabic}

\vspace{1.0cm}

\begin{center}
{\Large\sf Some issues in a gauge model of unparticles}
\\[10pt]
\vspace{.5 cm}

{Yi Liao\footnote{liaoy@nankai.edu.cn}}

{Department of Physics, Nankai University, Tianjin 300071, China}

\vspace{2.0ex}

{\bf Abstract}

\end{center}

We address in a recent gauge model of unparticles the issues that
are important for consistency of a gauge theory, i.e., unitarity
and Ward identity of physical amplitudes. We find that
non-integrable singularities arise in physical quantities like
cross section and decay rate from gauge interactions of
unparticles. We also show that Ward identity is violated due to
the lack of a dispersion relation for charged unparticles although
the Ward-Takahashi identity for general Green functions is
incorporated in the model. A previous observation that the
unparticle's (with scaling dimension $d$) contribution to the
gauge boson self-energy is a factor $(2-d)$ of the particle's has
been extended to the Green function of triple gauge bosons. This
$(2-d)$ rule may be generally true for any point Green functions
of gauge bosons. This implies that the model would be trivial even
as one that mimics certain dynamical effects on gauge bosons in
which unparticles serve as an interpolating field.

\begin{flushleft}
PACS: 12.90.+b, 14.80.-j, 11.55.Bq
%

Keywords: unparticle, gauge theory, unitarity, Ward identity
\end{flushleft}

\newpage

\section{Introduction}

As the era of the Large Hadron Collider is approaching, many new
theoretical ideas have been contemplated that could potentially be
tested there. One of radical suggestions is perhaps that of
unparticle by Georgi \cite{Georgi:2007ek}. Such an object is by
definition not a particle, but some stuff that follows scale
invariance, though it may well arise from certain high energy
scale theory of particles. The scale invariance makes a dispersion
relation generally not possible for an unparticle; instead, it
determines its kinematics in terms of a parameter, called scaling
dimension, of its corresponding field. The very nature of the
invariance also implies that the field is generically non-local.
The latter results in novel features not seen in the particle
world, for instance, non-trivial interference in the time-like
regime \cite{Georgi:2007si} (see also Ref \cite{Cheung:2007ue}),
one particle to one unparticle transitions \cite{Liao:2007bx}, and
non-integral power laws of long distance forces between particles
mediated by unparticles \cite{Liao:2007ic} (see also Ref
\cite{others}), etc.

Unparticles must interact with particles to be physically relevant
since we manipulate particles in experiments, and the interactions
can be systematically organized in effective field theory.
Although most studies, both phenomenological and theoretical, cope
with bosonic unparticles that couple as a standard model singlet
to particles \cite{Chen:2007qr}, it is hard to imagine that
unparticles must not carry the standard model charges. As a matter
of fact, the first gauge model of unparticles has been constructed
in Ref \cite{Cacciapaglia:2007jq}. In this circumstance, fermionic
unparticles
\cite{Luo:2007bq,He:2008rv,Basu:2008rd,Liao:2008tj,Cacciapaglia:2008ns}
can equally well couple to particles, and their phenomenology
could be even more interesting than their bosonic counterparts
\cite{Liao:2008tj}.

In this work we continue our theoretical investigation on the
gauge model of \cite{Cacciapaglia:2007jq} and address some issues
that have only been lightly touched upon in \cite{Liao:2007fv}. A
gauge model of unparticles must pass the standard consistency
criteria like unitarity and Ward identities for scattering
amplitudes. We make these checks and find the answer is negative
for both. This means that the model is not yet amenable to
computing physical amplitudes involving unparticles in the initial
or final state. In \cite{Liao:2007fv}, we also observed that the
scalar unparticle contribution to the complete (not just the
imaginary part of as shown in \cite{Cacciapaglia:2007jq}) gauge
boson self-energy is exactly $(2-d)$ times that of a scalar
particle in the same representation, where $d$ is the scaling
dimension of the unparticle. We extend this to the case of triple
gauge bosons. This seems to indicate that this $(2-d)$ rule is
generally true. In that case, the model of
\cite{Cacciapaglia:2007jq} would be naive even as one that mimics
certain dynamical effects on gauge bosons in which unparticles
serve as an interpolating field.

The paper is organized as follows. We describe briefly the gauge
model in the next section and catalog the Feynman rules to be
employed in later sections. The derivation of the rules is outlined
in the appendix. In section \ref{sec:rule} we show explicitly that
the $(2-d)$ rule holds true for the Green function of triple gauge
bosons. This is then followed in section \ref{sec:unitarity} by
comparative unitarity checks for ungauged and gauged unparticles
using the simplest two-point Green functions of particle fields.
Although the Green functions in the gauge model fulfil
Ward-Takahashi identities, we demonstrate in section \ref{sec:ward}
that physical amplitudes involving unparticles and physical gauge
bosons do not satisfy Ward identities. We conclude with some remarks
in the final section.

\section{Gauge model and Feynman rules}
\label{sec:model}

The scale symmetry of a scalar unparticle field of scaling
dimension $d$ demands its inverse propagator to be proportional to
$(-p^2-i\epsilon)^{2-d}$, with $p$ being its momentum
\cite{Georgi:2007si,Cheung:2007ue}. This is a non-integral power
for a general real number $d\ge 1$, and thus corresponds to a
non-local field. The non-locality makes the conventional minimal
gauging not work. Fortunately, a similar non-local problem was
successfully dealt with some years ago in the context of
reproducing low energy Goldstone dynamics from dynamical quarks in
QCD \cite{Terning:1991yt,Holdom:1992fn}. The lesson has been
recently applied to gauging unparticle fields in
\cite{Cacciapaglia:2007jq}.

A scalar unparticle multiplet $\calU$ may be coupled to gauge
fields $A_\mu^a$ via the Wilson line. The action is
\cite{Cacciapaglia:2007jq}
\begin{eqnarray}
S&=&\int d^4x~d^4y~\calU^{\dagger}(x)E(x-y)F(x,y),
\label{eq_action}\\
F(x,y)&=&P\exp\left[-igT^a\int_x^yA_{\mu}^a~dw^{\mu}\right]\calU(y),
\end{eqnarray}
where $P$ denotes path-ordering that effects on the group generators
$T^a$ in the unparticle representation, and $g$ is the gauge
coupling. $i^{-1}E(z)$ is the Fourier transform of the inverse
propagator:
\begin{eqnarray}
E(z)&=&\int\frac{d^4p}{(2\pi)^4}e^{-ip\cdot z}\Etilde(p),\\
\Etilde^{-1}(p)&\equiv&D(p)=\frac{A(d)}{2\sin(\pi
d)}\frac{1}{(-p^2-i\epsilon)^{2-d}},
\end{eqnarray}
with $A(d)$ a $d$-dependent constant not essential for our purpose.
Note that putting an infrared cut-off in the propagator does not
modify our subsequent conclusions.

The action (\ref{eq_action}) contains gauge interactions that are
quadratic in unparticle fields but involve gauge fields to an
arbitrary order. There is no obstacle to derive their Feynman rules
though the procedure rapidly becomes more and more involved as the
number of gauge fields increases. Some details of it are given in
the appendix. We list below the vertices up to three gauge fields
that will be required in later sections.

The Feynman rules for the vertices up to two gauge fields are
known in \cite{Cacciapaglia:2007jq}. The
$A_\mu^a\calU\calU^\dagger$ vertex is (with $ig$ to be attached on
both sides)
\begin{eqnarray}
\Gamma^a_\mu(-p-q,p;q)=T^a(2p+q)_\mu E_1(p;q),\label{eq_Gamma1}
\end{eqnarray}
where momenta before the semicolon are for unparticles and those
after it for gauge bosons, with all momenta being incoming. The
$A_\mu^aA_\nu^b\calU\calU^\dagger$ vertex is (with $ig^2$ to be
attached on both sides and differing by a factor of $i$ from
\cite{Cacciapaglia:2007jq})
\begin{eqnarray}
\Gamma^{ab}_{\mu\nu}(-p-q_{12},p;q_1,q_2)
&=&g_{\mu\nu}\{T^a,T^b\}E_1(p;q_{12})\nonumber\\
&+&%
T^aT^b(2p+q_2)_{\nu}(2p+q_2+q_{12})_{\mu}E_2(p;q_{12},q_2)\nonumber\\
&+&%
T^bT^a(2p+q_1)_{\mu}(2p+q_1+q_{12})_{\nu}E_2(p;q_{12},q_1),
\label{eq_Gamma2}
\end{eqnarray}
where $q_{12}=q_1+q_2$, etc, and the following notations are used,
\begin{eqnarray}
E_1(a;b)&=&\frac{\Etilde(a+b)-\Etilde(a)}{(a+b)^2-a^2},\nonumber\\
E_2(a;b_1,b_2)&=&\frac{E_1(a;b_1)-E_1(a;b_2)}{(a+b_1)^2-(a+b_2)^2},
\nonumber\\
E_3(a;b_1,b_2;c)&=&\frac{E_2(a;b_1,c)-E_2(a;b_2,c)}{(a+b_1)^2-(a+b_2)^2}.
\end{eqnarray}
The notations are slightly improved over those in
\cite{Liao:2007fv} to better display symmetry.

Finally, the $A_\alpha^aA_\beta^bA_\gamma^c\calU\calU^{\dagger}$
vertex is (with $ig^3$ to be attached on both sides)
\begin{eqnarray}
&&\Gamma^{abc}_{\alpha\beta\gamma}(-p-q_{123},p;q_1,q_2,q_3)\nonumber\\
&=&%
T^c\{T^a,T^b\}g_{\alpha\beta}(q_{123}+q_{12}+2p)_{\gamma}
E_2(p;q_{123},q_{12})\nonumber\\
&+&%
\{T^a,T^b\}T^cg_{\alpha\beta}(q_3+2p)_{\gamma}E_2(p;q_{123},q_3)\nonumber\\
&+&%
T^aT^bT^c
(q_{123}+q_{23}+2p)_{\alpha}(q_{23}+q_3+2p)_{\beta}(q_3+2p)_{\gamma}
E_3(p;q_{123},q_{23};q_3)\nonumber\\
&+&%
T^bT^aT^c
(q_{123}+q_{31}+2p)_{\beta}(q_{31}+q_3+2p)_{\alpha}(q_3+2p)_{\gamma}
E_3(p;q_{123},q_{31},q_3)\nonumber\\
&+&%
2~{\rm perms}. \label{eq_Gamma3}
\end{eqnarray}

\section{$(2-d)$ rule for triple gauge bosons}
\label{sec:rule}

When the action in (\ref{eq_action}) is exponentiated and
integrated in the path integral over the unparticle fields, we
obtain an effective action of the gauge fields. It is in this
sense that the gauge model discussed here is parallel to the
non-local chiral quark model in low energy hadronic physics
\cite{Terning:1991yt,Holdom:1992fn}. In the latter case,
integration over chiral quarks with a momentum-dependent dynamical
mass results in an effective action for the Goldstone bosons and
external sources or gauge fields. The chiral quark fields serve as
an interpolating field that mimics the strong dynamics of QCD as
manifested in the low energy constants in chiral Lagrangian. It
therefore sounds reasonable to expect that the unparticle fields
in the gauge model should at least play a similar role in the
context of certain new strong dynamics at a high energy scale.
\begin{center}
\begin{picture}(430,120)(0,0)

\SetOffset(40,10)%
\DashCArc(40,60)(25,0,360){3}\DashCArc(40,60)(24,0,360){2.5}
\Photon(40,85)(40,100){2}{3}%
\Photon(20,46)(5,35){2}{3}\Photon(75,35)(60,46){2}{3}%
\CArc(40,60)(20,250,286)\ArrowArc(40,60)(20,286,290)
\Text(40,105)[]{{\footnotesize$1$}}%
\Text(80,40)[l]{{\footnotesize$3$}}%
\Text(0,40)[r]{{\footnotesize$2$}}%
\Text(40,30)[]{{\footnotesize$p$}}
\Text(70,60)[l]{{\footnotesize$+~1$ more}}%
\Text(40,10)[]{$(1)$}

\SetOffset(170,10) %
\DashCArc(40,60)(25,0,360){3}\DashCArc(40,60)(24,0,360){2.5}
\Photon(40,85)(40,100){2}{3}%
\Photon(40,35)(20,20){2}{4}\Photon(40,35)(60,20){-2}{4}%
\Text(40,105)[]{{\footnotesize$1$}}%
\Text(70,60)[l]{{\footnotesize$+~2$ more}}%
\Text(65,70)[l]{{\footnotesize$p$}}
\Text(40,10)[]{$(2)$}

\SetOffset(290,20) %
\DashCArc(40,60)(25,0,360){3}\DashCArc(40,60)(24,0,360){2.5}
\Photon(40,35)(20,20){2}{4}\Photon(40,35)(60,20){-2}{4}%
\Photon(40,35)(40,15){2}{3.5}%
\Text(66,60)[l]{{\footnotesize$p$}}%
\Text(40,0)[]{$(3)$}

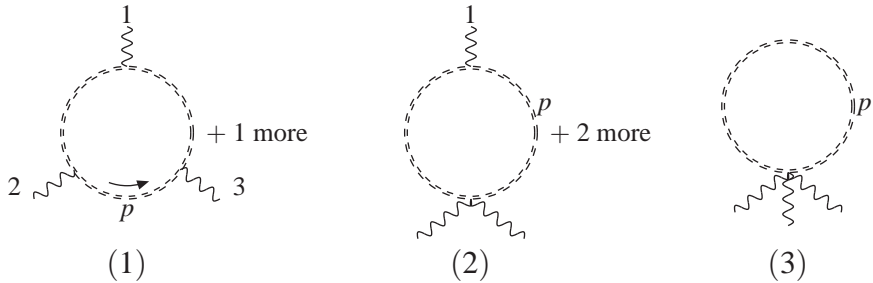
\SetOffset(0,10) \Text(200,-10)[]{Figure 1. Diagrams contributing
to triple gauge boson function.}

\end{picture}
\end{center}

In the previous work \cite{Liao:2007fv}, we observed that the
unparticle contribution to the gauge boson self-energy follows a
simple rule; namely, it is $(2-d)$ times the contribution from
scalar particles in the same representation. Its possible impact
on the running and unification of gauge couplings was also
studied. In this section we examine whether the rule applies to
the Green function of triple gauge bosons which is kinematically
more varied than a self-energy. If it does apply, it would
unlikely be accidental but might be generally true.

The contributing diagrams to the function are shown in Fig. 1, in
which the double dashed line stands for the scalar unparticle and
the three gauge bosons carry the group ($a,b,c$) and Lorentz
indices ($\alpha,\beta,\gamma$) with the incoming momenta $q_i$.
The vertices in section \ref{sec:model} yield
\begin{eqnarray}
\calA_{\alpha\beta\gamma}^{abc;1}&=&%
ig^3\tr\int\frac{d^4p}{(2\pi)^4}\Gamma^a_\alpha(-p+q_2,p+q_3;q_1)
\Gamma^c_\gamma(-p-q_3,p;q_3)\Gamma^b_\beta(-p,p-q_2;q_2)\nonumber\\
&\times&D(p)D(p+q_3)D(p-q_2)+(q_2,\beta,b)\leftrightarrow(q_3,\gamma,c),
\nonumber\\
\calA_{\alpha\beta\gamma}^{abc;2}&=&%
-ig^3\tr\int\frac{d^4p}{(2\pi)^4}\Gamma^a_\alpha(-p-q_1,p;q_1)
\Gamma^{bc}_{\beta\gamma}(-p,p+q_1;q_2,q_3)D(p)D(p+q_1)\nonumber\\
&+&2~{\rm perms},\nonumber\\
\calA_{\alpha\beta\gamma}^{abc;3}&=&%
ig^3\tr\int\frac{d^4p}{(2\pi)^4}\Gamma^{abc}_{\alpha\beta\gamma}
(-p,p;q_1,q_2,q_3)D(p). %
\label{eq_3point}
\end{eqnarray}
The integrals can be defined in $n$ dimensions for regularization,
but our subsequent algebraic manipulation does not depend on it.
We stress again that introducing an infrared cut-off to the
propagator does not modify our discussion either. The particle
case can be recovered in the limit $d\to 1$ whence $E_1\to 1$: The
graph (3) vanishes identically and the graph (2) vanishes due to
symmetry, while the two terms in graph (1) combine to
\begin{eqnarray}
\left[\calA_{\alpha\beta\gamma}^{abc}\right]_{\rm particle}
=g^3f^{abc}C(r)\int\frac{d^4p}{(2\pi)^4}\frac{(2p+q_3-q_2)_{\alpha}
(2p-q_2)_{\beta}(2p+q_3)_{\gamma}}
{[p^2+i\epsilon][(p+q_3)^2+i\epsilon][(p-q_2)^2+i\epsilon]},
\end{eqnarray}
where $\tr T^aT^b=C(r)\delta^{ab}$ for particles in representation
$r$.

In the unparticle case, the graph (3) contains two classes of
terms. The integrand in the first (class I) is proportional to
\begin{eqnarray}
E_1(p;0)=\lim_{q\to 0}E_1(p;q)
=\frac{2-d}{p^2+i\epsilon}\frac{1}{D(p)}, %
\label{eq_I}
\end{eqnarray}
while the remaining terms constitute the class II whose integrand
is not proportional to $E_1(p;0)$. We demonstrate that the class
II terms in graph (3) cancel completely the graphs (1) and (2).
First, the terms in class II that involve a single signature
tensor cancel in pair upon taking the traces and doing
integration, so do the similar terms in graph (2). Second, by
inspection, the remaining terms in II and graph (2) can be
combined in pair. Choosing judiciously the routing momenta and
making use of identities of $E_1$, we found that those terms
condense to a sum of two `form factors' with the common Lorentz
structure,
$(2p-q_2+q_3)^{\alpha}(2p+q_3)^{\gamma}(2p-q_2)^{\beta}$. This is
the same Lorentz structure exactly for the first term in
$\calA_{\alpha\beta\gamma}^{abc;1}$ and up to a minus sign for the
second upon flipping the sign of $p$. The first form factor is
(with the prefactors $ig^3$ understood),
\begin{eqnarray}
\tr T^aT^bT^c \sum_{(ijk)}\frac{d_j}{n_j-n_i}\left[d_ke_i(e_i-e_j)
+\frac{e_i}{n_j-n_k}+\frac{e_k-e_i}{n_i-n_j}\right],
\label{eq_fraction}
\end{eqnarray}
where the sum is over the set $(123),(231),(312)$. We have
introduced some abbreviations:
\begin{eqnarray}
&&n_1=p^2,~~n_2=(p-q_2)^2,~~n_3=(p+q_3)^2;\nonumber\\
&&d_1=D(p),~~d_2=D(p-q_2),~~d_3=D(p+q_3);\nonumber\\
&&e_i=\frac{d^{-1}_j-d^{-1}_k}{n_j-n_k},
\end{eqnarray}
where $ijk$ is again cyclic in $123$. The second form factor is
obtained from the first by the interchanges of indices,
$b\leftrightarrow c$ and $2\leftrightarrow 3$, but with a global
minus sign. Third, using identities of fractions, it is
straightforward to show that the sum in eq (\ref{eq_fraction}) is
equal to $d_1d_2d_3e_1e_2e_3$. This is again the integrand in the
second term of $\calA_{\alpha\beta\gamma}^{abc;1}$ upon extracting
the Lorentz structure displayed above so that the sum in
(\ref{eq_fraction}) (i.e., the first form factor) completely
cancels the second term in $\calA_{\alpha\beta\gamma}^{abc;1}$.
Similar cancellation occurs between the second form factor and the
first term in eq (\ref{eq_fraction}). To summarize, the complete
graphs (1) and (2) are cancelled by class II terms in graph (3).

Now we are left with class I terms in graph (3). First of all, eqs
(\ref{eq_I},\ref{eq_3point}) imply that the integrand is of a
particle type. The terms with a signature tensor are again
cancelled in pair. Using the fraction identity,
$\displaystyle\sum_{(ijk)}[n_i(n_i-n_j)(n_i-n_k)]^{-1}=[n_1n_2n_3]^{-1}$,
some algebra verifies our claim:
\begin{eqnarray}
\left[\calA_{\alpha\beta\gamma}^{abc}\right]_{\rm unparticle}
=(2-d)\left[\calA_{\alpha\beta\gamma}^{abc}\right]_{\rm particle}.
\end{eqnarray}

We end this section with a remark. It seems unlikely that the
above relation is specific to two- and three-point functions of
gauge fields. Our explicit demonstration of it might suggest a way
to reach the general result for any point functions: The
cancellation mechanism witnessed in two- and three-point functions
might indicate that the only contribution for any point function
comes exclusively from the tad-pole like graph involving the
highest point vertex in each case.

\section{Unitarity}
\label{sec:unitarity}

Unitarity of the scattering matrix is one of the fundamental
criteria that any quantum theory must meet. This is especially
true of a gauge theory in which additional delicacies may occur.
The purpose of this section is to show using the simplest possible
process that the gauge model of unparticles proposed in
\cite{Cacciapaglia:2007jq} breaks unitarity. For comparison, we
also examine unitarity in non-gauge interactions between
unparticles and particles, and we find that these interactions
generally preserve unitarity in the conventional sense.

Consider the one-loop self-energy of a scalar particle field arising
from interactions with scalar unparticles. To the graphs (1) and (2)
shown in Fig. 2 there correspond the two effective interactions:
\begin{eqnarray}
\calL_1&=&\lambda_1\Phi\varphi\calU,\label{eq_L1}\\
\calL_2&=&\lambda_2\Phi\calU_1\calU_2,\label{eq_L2}
\end{eqnarray}
where $\Phi,~\varphi$ are the scalar particle fields of mass
$M,~m$, and $\calU,~\calU_1,~\calU_2$ the scalar unparticle fields
of scaling dimension $d,~d_1,~d_2$, respectively.
\begin{center}
\begin{picture}(240,100)(0,0)
\SetOffset(0,30) %
\DashLine(10,40)(25,40){3}\DashLine(75,40)(90,40){3}%
\DashCArc(50,40)(25,0,360){2}\DashCArc(50,40)(24,0,180){2} %
\Text(5,40)[]{{\small$p$}}\Text(50,58)[]{{\small$k$}}
\Text(50,0)[]{(1)}

\SetOffset(120,30) %
\DashLine(10,40)(25,40){3}\DashLine(75,40)(90,40){3}%
\DashCArc(50,40)(25,0,360){2}\DashCArc(50,40)(24,0,360){2} %
\Text(5,40)[]{{\small$p$}}\Text(50,58)[]{{\small$k$}}
\Text(50,0)[]{(2)}

\SetOffset(0,10) %
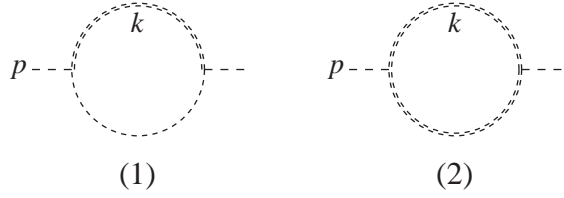
\Text(100,0)[]{Figure 2. Self-energy of scalar particle field
arising from eqs (\ref{eq_L1},\ref{eq_L2}).}
\end{picture}
\end{center}

The imaginary part of the self-energy in graph (2) is found to be,
for $M>m$,
\begin{eqnarray}
\textrm{Im}~\calA_1(M^2)&=&
\frac{\lambda_1^2M^{2(d-1)}}{(4\pi)^2}\frac{A(d)}{2(d-1)}
\int_{r^2}^1dx~x^{1-d}(1-x)^{d-1}(x-r^2)^{d-1}\nonumber\\
&=& \frac{\lambda_1^2M^{2(d-1)}}{(4\pi)^2}\frac{A(d)}{2(d-1)}
(1-r^2)^{2d-1}~_2F_1(d-1,d;2d;1-r^2)B(d,d),
\end{eqnarray}
where $r=m/M$ and $_2F_1$ and $B$ are the standard special
functions. This should be compared to the decay width for
$\Phi\to\varphi+\calU$ for unitarity check. Finishing all phase
space integrals but that of the unparticle energy yields
\begin{eqnarray}
\Gamma_1=\frac{\lambda_1^2M^{2(d-1)}}{(2\pi)^2}\frac{A(d)}{2^{3-d}M}
(1-r)^d \int_{\frac{1}{2}(1+r)}^1
dt~\left[(1-t)\left(\frac{1+r}{1-r}-t\right)\right]^{\frac{1}{2}}
\left[t-\frac{1}{2}(1+r)\right]^{d-2}.
\end{eqnarray}
Changing the variable to $u=[t-\frac{1}{2}(1+r)]/[\frac{1}{2}(1-r)]$
works out the integral to
\begin{eqnarray}
\Gamma_1&=&\frac{\lambda_1^2M^{2(d-1)}}{(4\pi)^2}\frac{A(d)}{2M}
(1-r)^{2d-1}(1+r)\nonumber\\
&\times&_2F_1\left(-\frac{1}{2},d-1;d+\frac{1}{2};
\left(\frac{1-r}{1+r}\right)^2\right)
B\left(d-1,\frac{3}{2}\right).
\end{eqnarray}
The unitarity relation $\textrm{Im}~\calA_1(M^2)=M\Gamma_1$ is
verified using the relation
\begin{eqnarray}
F\left(d-1,d;2d;\frac{4z}{(1+z)^2}\right)
=2^{2(d-1)}(1+r)^{-2(d-1)}F\left(-\frac{1}{2},d-1;d+\frac{1}{2};z^2\right),
\end{eqnarray}
and a relation for $B$ function.

The interaction $\calL_2$ involves two unparticles and has been
less discussed in the literature. Its contribution to the
imaginary part of graph (2) is easily worked out to be
\begin{eqnarray}
\textrm{Im}~\calA_2(M^2)&=&-\frac{\lambda_2^2A(d_1)A(d_2)}{4(4\pi)^2}
M^{2(d_1+d_2-2)}\nonumber\\
&\times&\frac{\Gamma(2-d_1-d_2)}{\Gamma(2-d_1)\Gamma(2-d_2)}
B(d_1,d_2)\frac{\sin(d_1+d_2)\pi}{\sin(d_1\pi)\sin(d_2\pi)}.
\end{eqnarray}
The phase space for the decay $\Phi\to\calU_1\calU_2$ is more
involved. Finishing integrals of one unparticle momentum and the
angles of the other, we obtain
\begin{eqnarray}
\Gamma_2=
\frac{\lambda_2^2}{(2\pi)^3M}A(d_1)A(d_2)M^{2(d_1+d_2-2)}I_{d_1-2,d_2-2},
\end{eqnarray}
where
\begin{eqnarray}
I_{\alpha,\beta}&=&\iint_{R}
dv_0dv~v^2(v_0^2-v^2)^\alpha[(1-v_0)^2-v^2]^\beta.
\end{eqnarray}
Here the integration region $R$ in the $vv_0$ plane is bounded by
the lines, $v=0$, $v=v_0$, and $v+v_0=1$. As the integrand is even
in $v$, we make the region symmetric under $v\to -v$. The integrals
are then factorized by the new variables, $v_0-v=x$, $v_0+v=y$ with
$x\in[0,1]$ and $y\in[0,1]$ so that, for $\alpha>-1,~\beta>-1$
(i.e., $d_{1,2}>1$ in our case),
\begin{eqnarray}
I_{\alpha,\beta}&=&2^{-4}\int_0^1dx\int_0^1dy~
(y-x)^2(xy)^\alpha[(1-y)(1-x)]^\beta\nonumber\\
&=&2^{-3}\left[B(\alpha+3,\beta+1)B(\alpha+1,\beta+1)
-\left(B(\alpha+2,\beta+1)\right)^2\right]\nonumber\\
&=&\frac{\Gamma(\alpha+1)\Gamma(\alpha+2)\Gamma(\beta+1)\Gamma(\beta+2)}
{8\Gamma(\alpha+\beta+3)\Gamma(\alpha+\beta+4)}.
\end{eqnarray}
The unitarity relation $M\Gamma_2=\textrm{Im}~\calA_2(M^2)$ is
confirmed using $\Gamma(z)\Gamma(1-z)\sin(z\pi)=\pi$.

It is not surprising that unitarity is preserved by non-gauge
interactions of unparticles because the unparticle propagator has
the correct cut structure by construction \cite{Georgi:2007si} and
because those interactions are Hermitian. As we pointed out in
\cite{Liao:2007fv}, the gauge interactions of unparticles in
\cite{Cacciapaglia:2007jq} are actually non-Hermitian in the
time-like regime. This may be a source of unitarity violation in
the model. In Ref \cite{Liao:2007fv}, we reached this conclusion
by symmetry analysis for the process, $q\bar q\to\calU\bar\calU$
via gluon exchange assuming $\calU$ is charged under QCD. In what
follows, we demonstrate the violation analytically by the simplest
possible process of the gauge boson decay,
$A_\mu^a(p)\to\calU(k_1)\bar\calU(k_2)$, with the gauge boson
momentum $p$ in the time-like regime.

The imaginary part of the gauge boson self-energy from the
unparticle loop can easily be obtained from that of the scalar
particle's using the $(2-d)$ rule:
\begin{eqnarray}
\Pi^{ab}_{\mu\nu}(p)&=&\delta^{ab} \left(\frac{p_\mu
p_\nu}{p^2}-g_{\mu\nu}\right)\Pi(p^2),\nonumber\\
\text{Im~}\Pi(p^2)&=&(2-d)\frac{g^2}{48\pi}C(r)p^2.
\end{eqnarray}
The amplitude for the decay is basically the vertex shown in eq
(\ref{eq_Gamma1}). The properly summed and averaged decay rate is
\begin{eqnarray}
\Gamma=\frac{2}{3\pi^3}g^2C(r)\sqrt{p^2} \sin^2(d\pi)J(d),
\end{eqnarray}
where
\begin{eqnarray}
J(d)=\iint_R dv_0dv\frac{v^4}{(1-2v_0)^2}
\left[\left(\frac{v_0^2-v^2}{(1-v_0)^2-v^2}\right)^{2-d}
+\left(\frac{(1-v_0)^2-v^2}{v_0^2-v^2}\right)^{2-d}-2\right],
\end{eqnarray}
with the same region $R$ as in $\Gamma_2$ above. The terms in the
square brackets arise from combination of phase space factors and
the numerator of $E_1$ in the vertex. The potential singularity at
$v_0=\frac{1}{2}$ is due to the denominator in $E_1$, and makes it
impossible to factorize the integral into two separate ones as we
did in $I_{\alpha,\beta}$.

To observe the non-integrability that the singularity may cause,
we finish the $v$ integral first. Using the symmetry with respect
to $v_0=\frac{1}{2}$, we restrict ourselves to the left half of
$R$ and obtain
\begin{eqnarray}
J(d)&=&\frac{1}{2}\int_0^1dt\frac{t^2}{(1+\sqrt{t})^3(1-t)^2}f(d-2;t),
\end{eqnarray}
where
\begin{eqnarray}
f(a;t)&=&t^a B\left(\frac{5}{2},1+a\right)
~_2F_1\left(a,\frac{5}{2};\frac{7}{2}+a;t\right)\nonumber\\
&+&t^{-a}B\left(\frac{5}{2},1-a\right)
~_2F_1\left(-a,\frac{5}{2};\frac{7}{2}-a;t\right) -\frac{4}{5}.
\end{eqnarray}
Note that $J(d)$ is even in $(2-d)$ (so is $\Gamma$) while
$\text{Im~}\Pi(p^2)$ is odd. This is the argument employed in
\cite{Liao:2007fv} to signify unitarity breakdown. But what really
occurs is even worse: The singularity introduced by the vertex in eq
(\ref{eq_Gamma1}) is logarithmically non-integrable. To see this,
one needs to expand $f(a;1-z)$ at $z=0$:
\begin{eqnarray}
f(a;1-z)&=&a\left[\psi(1+a)-\psi(1-a)\right]z\nonumber\\
&+&\left(\frac{3}{2}a^2\ln z+
a^2\left[C-\psi(1-a)-\psi(1+a)\right]\right.\nonumber\\
&&\left.-\frac{7}{4}a(1-a)\psi(2-a)
+\frac{7}{4}a(a+1)\psi(2+a)\right)z^2+O(z^3),
\end{eqnarray}
where $\psi(\xi)=\Gamma'(\xi)/\Gamma(\xi)$ and $C$ is a constant.
The unitarity is thus badly violated in this case by
non-integrable singularities introduced in interaction vertices of
the gauge model. Note in passing that there is no problem with the
particle limit of $d\to 1$ although it is better to take the limit
at the very start to avoid the ambiguity between the
$\sin^2(d\pi)$ factor in $\Gamma$ and the singularity. We stress
that this breakdown of unitarity occurs at all energy scales in
the gauge model of unparticles, in contrast to the conventional
effective field theory in which unitarity starts to be violated at
energy scales close to its ultraviolet cut-off. This also implies
that unitarity cannot be simply recovered by including new degrees
of freedom in the gauge model as we do in a conventional effective
theory.

\section{Ward identity}
\label{sec:ward}

Now we address the issue of Ward identity necessary for a consistent
gauge theory. We do not expect any problem with Ward-Takahashi
identity for generally off-shell Green functions as it is built in
by construction of the gauge model. As examples, we list below the
first few identities for Green functions involving up to three gauge
bosons.

The simplest one is
\begin{eqnarray}
q^{\mu}\Gamma^a_\mu(-p-q,p;q)=T^a[D^{-1}(p+q)-D^{-1}(p)],
\end{eqnarray}
while the next one requires a little rearrangement of terms,
\begin{eqnarray}
&&q_1^\mu\Gamma^{ab}_{\mu\nu}(-p-q_{12},p;q_1,q_2)\nonumber\\
&=&\Gamma^b_\nu(-p-q_{12},p+q_1;q_2)T^a-T^a\Gamma^b_\nu(-p-q_2,p;q_2)
\nonumber\\
&+&if^{abc}\Gamma^c_\nu(-p-q_{12},p;q_{12}).
\end{eqnarray}
The derivation of identity for the triple gauge boson vertex is much
more involved. The main trick is to use partial fraction. But when
the cloud clears up, the answer is simple:
\begin{eqnarray}
&&q_1^\alpha\Gamma^{abc}_{\alpha\beta\gamma}(-p-q_{123},p;q_1,q_2,q_3)
\nonumber\\
&=&%
\Gamma^{bc}_{\beta\gamma}(-p-q_{123},p+q_1;q_2,q_3)T^a-
T^a\Gamma^{bc}_{\beta\gamma}(-p-q_{23},p;q_2,q_3)\nonumber\\
&+&%
if^{abd}\Gamma^{dc}_{\beta\gamma}(-p-q_{123},p;q_{12},q_3)-
if^{cad}\Gamma^{bd}_{\beta\gamma}(-p-q_{123},p;q_2,q_{31}).
\end{eqnarray}

In the conventional gauge theory of particles, the Ward identity for
physical amplitudes is obtained from the Ward-Takahashi identity by
going to the physical limit of charged particles. In an Abelian
theory like QED it is sufficient to require electrons to be
on-shell. But in a non-Abelian theory like QCD, it is necessary that
gluons be physical as well since they are also charged. In a gauge
theory of unparticles however, an on-shell condition (dispersion
relation) is missing for unparticles; this may endanger the Ward
identity for physical amplitudes. If this happens, unphysical states
of gauge bosons can be produced by unparticles, which is of course
not acceptable. We show below by a simple process that this happens
indeed in the considered model.

Consider the process of unparticle pair production by the fusion of
a gauge boson pair,
$A^a_\alpha(k_1)A^b_\beta(k_2)\to\calU(p_1)\bar\calU(p_2)$, whose
Feynman diagrams are depicted in Fig. 3. Putting the gauge bosons
on-shell, $k_1^2=k_2^2=0$, the once contracted component amplitudes
are
\begin{eqnarray}
k_1^{\alpha}\calA^{ab;1}_{\alpha\beta}&=&
-[T^a,T^b]\left[k_{1\beta}(p_1^2-p_2^2)+k_{2\beta}k_1\cdot(p_1-p_2)
+(p_2-p_1)_{\beta}2k_1\cdot k_2\right]\nonumber\\
&\times&\frac{E_1(-p_1;p_1+p_2)}{2k_1\cdot k_2},\nonumber\\
k_1^{\alpha}\calA^{ab;2}_{\alpha\beta}&=&
\{T^a,T^b\}k_{1\beta}E_1(-p_1;p_1+p_2)\nonumber\\
&+&T^aT^b(-2p_1+k_2)_{\beta}\left[E_1(-p_1;p_1+p_2)-E_1(-p_1;k_2)\right]
\nonumber\\
&+&T^bT^a\frac{-k_1\cdot p_1}{k_2\cdot
p_2}(-2p_1+2k_1+k_2)_{\beta}\left[E_1(-p_1;p_1+p_2)-E_1(-p_1;k_1)\right],
\nonumber\\
k_1^{\alpha}\calA^{ab;3}_{\alpha\beta}&=&
-T^bT^a(2p_2-k_2)_{\beta}\frac{k_1\cdot p_1}{k_2\cdot p_2}
\left[1-\frac{D(p_2-k_2)}{D(p_2)}\right]E_1(-p_1;k_1),\nonumber\\
k_1^{\alpha}\calA^{ab;4}_{\alpha\beta}&=&
T^aT^b\left[1-\frac{D(p_2-k_1)}{D(p_2)}\right](-2p_1+k_2)_{\beta}E_1(-p_1;k_2),
\end{eqnarray}
where a $g^2$ factor is implied on the right hand side. The particle
case is recovered by sending $D^{-1}(q)\to q^2$ and $E_1\to 1$. As
the above result seems hopeless, we examine the simpler, less
restrictive, doubly contracted amplitude. The sum is
\begin{eqnarray}
k_1^{\alpha}k_2^{\beta}\calA^{ab}_{\alpha\beta}&=&
\frac{T^aT^b}{D(p_1)D(p_2)}
\left(D(k_1-p_2)-\frac{1}{2}\left[D(p_1)+D(p_2)\right]\right)\nonumber\\
&+&\frac{T^bT^a}{D(p_1)D(p_2)}
\left(D(k_2-p_2)-\frac{1}{2}\left[D(p_1)+D(p_2)\right]\right),
\end{eqnarray}
which does not vanish as the Ward identity requires. The situation
does not improve in the Abelian case.

\begin{center}
\begin{picture}(440,110)(0,0)

\SetOffset(0,30) %
\Photon(10,10)(40,40){2}{5}\Photon(40,40)(10,70){2}{5}\Photon(40,40)(80,40){2}{5}
\DashLine(109,10)(79,40){3}\DashLine(79,40)(109,70){3}%
\DashLine(111,10)(81,40){3}\DashLine(81,40)(111,70){3}%
\Text(10,5)[l]{$1$}%
\Text(10,75)[l]{$2$}%
\Text(110,5)[r]{$1$}\Text(110,75)[r]{$2$}%
\Text(60,-10)[]{$(1)$}

\SetOffset(130,30) %
\Photon(10,10)(40,40){2}{5}\Photon(40,40)(10,70){2}{5}%
\DashLine(69,10)(39,40){3}\DashLine(39,40)(69,70){3}
\DashLine(71,10)(41,40){3}\DashLine(41,40)(71,70){3}
\Text(40,-10)[]{$(2)$}

\SetOffset(220,30) %
\Photon(10,10)(50,10){2}{5}\Photon(10,70)(50,70){2}{5}
\DashLine(90,9)(49,9){3}\DashLine(49,9)(49,71){3}\DashLine(49,71)(90,71){3}
\DashLine(90,11)(51,11){3}\DashLine(51,11)(51,69){3}\DashLine(51,69)(90,69){3}
\Text(50,-10)[]{$(3)$}

\SetOffset(330,30) %
\Photon(10,10)(50,70){2}{9}\Photon(10,70)(50,10){-2}{9}
\DashLine(90,9)(49,9){3}\DashLine(49,9)(49,71){3}\DashLine(49,71)(90,71){3}
\DashLine(90,11)(51,11){3}\DashLine(51,11)(51,69){3}\DashLine(51,69)(90,69){3}
\Text(50,-10)[]{$(4)$}

\SetOffset(0,0)%
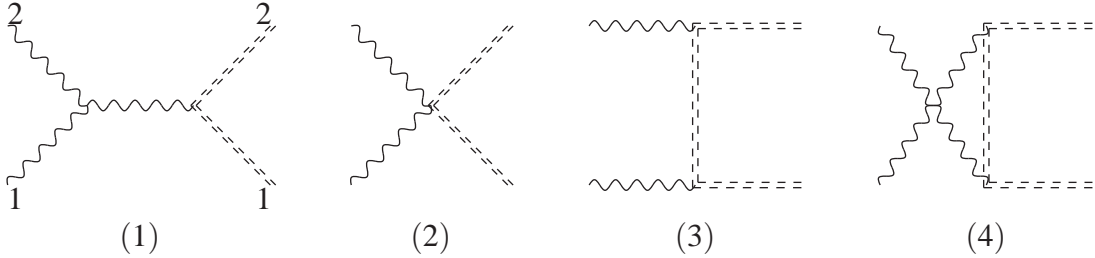
\Text(210,0)[]{Figure 3. Diagrams contributing to the process
$A^a_\alpha(k_1)A^b_\beta(k_2)\to\calU(p_1)\bar\calU(p_2)$.}
\end{picture}
\end{center}

A simple way out seems to require $D^{-1}(p)=0$ for unparticles
appearing in the initial and final states. While the meaning of
this is obscure by itself, it implies a dispersion relation for
unparticles with $d<2$. But this is obviously not a consistent
concept as there would be no difference between a physical
particle and unparticle. Furthermore, it would break unitarity
established for unparticles that have no gauge interactions.

\section{Conclusion}

It looks natural that unparticles are charged under the standard
model gauge group. But it is rather difficult to couple them to
gauge fields since they are generically nonlocal in nature.
Nevertheless, the first gauge model has been attempted in Ref
\cite{Cacciapaglia:2007jq} (for recent discussions about the work
and development, see
\cite{Licht:2008ic,Galloway:2008jn,Ilderton:2008ab}). We have
investigated in this work whether the model fulfils the standard
requirements that a consistent gauge theory must do; namely, the
unitarity and the Ward identity for physical amplitudes involving
unparticles. We find that the answers to both are negative.

In non-gauge interactions no surprise is expected for unitarity as
the unparticle propagator incorporates correct analyticity
properties and the interactions are usually trivial in analyticity
structure. The latter is no longer the case in the gauge model
considered here. We find that its gauge interactions introduce
non-integrable singularities in phase space that make physical
quantities like cross section meaningless. This failure in
unitarity occurs at any energy scale and thus cannot be cured by
incorporating new degrees of freedom as one does in a conventional
effective field theory. The result on Ward identity may be
surprising at first sight since the Ward-Takahashi identity for
general Green functions has been built in the model. The point
here is that to obtain the Ward identity for physical amplitudes
some physical conditions have to be imposed to delete the contact
terms in the Ward-Takahashi identity. These are the on-shell
condition for charged particles and the transversality condition
for gauge bosons if charged. It is the lack of a dispersion
relation for charged unparticles that the passage is blocked.
These two defects might easily be blamed upon the wisdom of
conformal field theory that no consistent scattering matrix is
known. But this does not explain why we seem to obtain reasonable
results with non-gauge interactions of unparticles by following
the standard procedure in field theory.

Even if a gauge model of unparticles is afflicted with these
diseases, it could still serve as a useful tool to mimic certain
strong dynamical effects on gauge bosons at low energies. In that
case, unparticles appear as an interpolating field confined to the
virtual loops of gauge bosons. This situation is parallel to the
relationship between the nonlocal chiral quark model
\cite{Terning:1991yt,Holdom:1992fn} and the low energy dynamics of
Goldstone bosons. Because of this, we have considered the Green
function of triple gauge bosons due to unparticle loops and found
that an earlier observation on the self-energy of gauge bosons
also applies here. Namely, the unparticle contribution to the
Green function is a factor $(2-d)$ of the scalar particle's in the
same representation of the gauge group. We conjecture that this
$(2-d)$ rule may be generally true. If this were the case, the
model would be too naive even if the unparticles are considered as
an interpolating field. It looks fair to say that the challenge of
gauging unparticles still remains.

\vspace{0.5cm}
\noindent %
{\bf Acknowledgement} This work is supported in part by the grants
NCET-06-0211 and NSFC-10775074.

\vspace{0.5cm}
\noindent %
{\bf Appendix. Derivation of Feynman rules}

We outline the derivation of the vertices in eqs
(\ref{eq_Gamma1},\ref{eq_Gamma2},\ref{eq_Gamma3}) for
completeness. The basic technique was developed in Ref
\cite{Terning:1991yt}. The vertices with up to two gauge bosons
were known previously \cite{Terning:1991yt,Cacciapaglia:2007jq}
for which our manipulation of series is slightly simpler, while
the vertex with three gauge bosons is not yet available in the
literature. Introducing Fourier transforms of the functions
\begin{eqnarray}
\calU(x)&=&\int(dk)e^{-ik\cdot x}\Utilde(k),\nonumber\\
F(x,y)&=&\int(dq_1)(dq_2)e^{-i(q_1\cdot x+q_2\cdot
y)}\Ftilde(q_1,q_2),
\end{eqnarray}
and assuming $\Etilde(p)$ is a function of $p^2$ (which is the
case here), expansion at $p^2=0$ yields for the action in eq
(\ref{eq_action}):
\begin{eqnarray}
S&=&\sum_{n=0}\frac{\Etilde^{(n)}(0)}{n!}\int(dk)(dp)
\Utilde^{\dagger}(k)\left(p^2\right)^n\Ftilde(k-p,p)\nonumber\\
&=&\sum_{n=0}\frac{\Etilde^{(n)}(0)}{n!}\int(dk)(dp)\int
dxdy~e^{i[-k\cdot x+(k-p)\cdot z+p\cdot y]}
\calU^{\dagger}(x)\left(p^2\right)^nF(z,y).
\end{eqnarray}
Replacing $e^{ip\cdot y}(p^2)^n=[(-\partial^2)^ne^{ip\cdot y}]$ and
doing integration by parts, the action becomes
\begin{eqnarray}
S=\sum_{n=0}\frac{\Etilde^{(n)}(0)}{n!}\int dxdy~
\delta(x-y)\calU^{\dagger}(x)\left(-\partial^2\right)^n\left[
P\exp\left(-igT^a\int_x^yA_{\mu}^a~dw^{\mu}\right)\calU(y)\right],
\end{eqnarray}
which is the starting point to all vertices. We have used the
following abbreviations:
\begin{eqnarray}
(dk)=\frac{d^4k}{(2\pi)^4},~dx=d^4x,~\delta(x)=\delta^4(x),
\end{eqnarray}
and the derivatives always refer to $y$ unless otherwise stated.

We begin with the derivation of eq (\ref{eq_Gamma1}). The vertex in
coordinate space is
\begin{eqnarray}
&&\left.\frac{\delta^3S}{\delta
A_{\mu}^a(x_1)\delta\calU(v)\delta\calU^{\dagger}(z)}\right|_0
\nonumber\\
&=&\sum_{n=0}\frac{\Etilde^{(n)}(0)}{n!}\int dy~
\delta(y-z)\left(-\partial^2\right)^n\left[
(-ig)L_1^{a\mu}\delta(y-v)\right], \label{eq_V1}
\end{eqnarray}
where from now on the following short-cuts will be used,
\begin{eqnarray}
L^{a\alpha}_i=T^a\int_z^y\delta(x_i-u)du^{\alpha},
~\delta_i=\delta(x_i-y).
\end{eqnarray}
Its Fourier transform yields
\begin{eqnarray}
&&\int dx_1dvdz~e^{i[p'z-pv-qx_1]}(\ref{eq_V1})\nonumber\\
&=&\sum_{n=0}\frac{\Etilde^{(n)}(0)}{n!}\int dx_1dydz~
e^{i[p'z-py-qx_1]}\left[\left(-\partial^2\right)^n\delta(z-y)\right]
\left[-igL^{a\mu}_1\right].
\end{eqnarray}
The basic trick here is integration by parts. The $n=0$ term
vanishes, while the $n=1$ term gives, using
$\partial^{\nu}L_1^{a\mu}=T^a\delta_1g^{\mu\nu}$,
\begin{eqnarray}
(n=1~\text{term})=\Etilde^{(1)}(0)gT^a(p+p')^{\mu}~(2\pi)^4\delta(p'-p-q).
\end{eqnarray}
The general term is obtained by induction. Assuming
\begin{eqnarray}
(n{\rm -th~term})=
\frac{\Etilde^{(n)}(0)}{n!}gT^a(p+p')^{\mu}f_n~(2\pi)^4\delta(p'-p-q),
\end{eqnarray}
with $f_0=0,~f_1=1$, the coefficient for the next term is found to
be
\begin{eqnarray}
f_{n+1}=p^2f_n+(p+q)^{2n},
\end{eqnarray}
with $k^{2n}\equiv(k^2)^n$; namely,
\begin{eqnarray}
\frac{f_{n+1}}{(p+q)^{2n}}=1+r\frac{f_n}{(p+q)^{2(n-1)}}
=1+r+\cdots+r^n=\frac{1-r^{n+1}}{1-r},~~~r\equiv\frac{p^2}{(p+q)^2}.
\end{eqnarray}
The general coefficient is thus
\begin{eqnarray}
f_n=\frac{(p+q)^{2n}-p^{2n}}{(p+q)^2-p^2}.
\end{eqnarray}
Using
\begin{eqnarray}
\sum_n\frac{\Etilde^{(n)}(0)}{n!}\frac{(p+q)^{2n}-p^{2n}}{(p+q)^2-p^2}
=\frac{\Etilde(p+q)-\Etilde(p)}{(p+q)^2-p^2},
\end{eqnarray}
eq (\ref{eq_Gamma1}) is obtained.

Now we derive the vertex in eq (\ref{eq_Gamma2}):
\begin{eqnarray}
&&\int dx_1dx_2dvdz~e^{i[p'z-pv-q_1x_1-q_2x_2]}
\left.\frac{\delta^4S}{\delta A_{\mu}^a(x_1)\delta A_{\nu}^b(x_2)
\delta\calU(v)\delta\calU^{\dagger}(z)}\right|_0\nonumber\\
&=&-g^2\sum_{n=0}\frac{\Etilde^{(n)}(0)}{n!}\int
dx_1dx_2dydz~e^{i[p'z-py-\sum q_ix_i]}
\left[\left(-\partial^2\right)^n\delta(y-z)\right]
P\left[L_1^{a\mu}L_2^{b\nu}\right].
\end{eqnarray}
The $n=0$ term again vanishes, while the $n=1$ term gives
\begin{eqnarray*}
(n=1){\rm ~term}
=g^2\Etilde^{(1)}(0)g^{\mu\nu}\{T^a,T^b\}~(2\pi)^4\delta(p'-p-q_{12}),
\end{eqnarray*}
with $q_{12}=q_1+q_2$. Since not all possible Lorentz structures
have appeared, we have to go one step further. The $n=2$ term
yields after some algebra
\begin{eqnarray}
(n=2){\rm ~term}&=&
g^2\frac{\Etilde^{(2)}(0)}{2!}(2\pi)^4\delta(p'-p-q_{12})
\left\{g^{\mu\nu}\{T^a,T^b\}(p^{\prime 2}+p^2)\right.\nonumber\\
&&+T^bT^a(2p+q_1)^{\mu}(2p+q_1+q_{12})^{\nu}\nonumber\\
&&\left.+T^aT^b(2p+q_2)^{\nu}(2p+q_2+q_{12})^{\mu}\right\}.
\end{eqnarray}
Assuming the coefficients for the $\{T^a,T^b\}$, $T^bT^a$ and
$T^aT^b$ terms to be $g_n(p,q_1,q_2)$, $h_n(p,q_1,q_2)$ and
$h_n(p,q_2,q_1)$ respectively, with the initial conditions:
\begin{eqnarray}
g_0=0,~g_1=1,~g_2=p^{\prime 2}+p^2;~h_0=h_1=0,~h_2=1,
\end{eqnarray}
we find out their $(n+1)$-th expressions after some work:
\begin{eqnarray}
g_{n+1}(p,q_1,q_2)&=&p^2g_n(p,q_1,q_2)+(p+q_{12})^{2n},
\nonumber\\
h_{n+1}(p,q_2,q_1)&=&p^2h_n(p,q_2,q_1)+
\frac{(p+q_{12})^{2n}-(p+q_2)^{2n}}{(p+q_{12})^{2}-(p+q_2)^2}.
\end{eqnarray}
$g_n$ has the same structure as $f_n$ while $h_n$, when multiplied
by $[(p+q_{12})^{2}-(p+q_2)^2]$, becomes a difference of two
series each having the same structure as $f_n$ again. We thus find
\begin{eqnarray}
g_n(p,q_1,q_2)&=&\frac{(p+q_{12})^{2n}-p^{2n}}{(p+q_{12})^2-p^2},
\nonumber\\
h_n(p,q_2,q_1)&=&\frac{1}{(p+q_{12})^{2}-(p+q_2)^2}\left[
\frac{(p+q_{12})^{2n}-p^{2n}}{(p+q_{12})^2-p^2}-
\frac{(p+q_2)^{2n}-p^{2n}}{(p+q_2)^2-p^2}\right].
\end{eqnarray}
Then eq (\ref{eq_Gamma2}) obtains readily.

Finally, we describe briefly the derivation of the vertex with three
gauge bosons. The algebra blows up rapidly as the number of gauge
bosons increases. The vertex to compute is
\begin{eqnarray}
&&\int\Pi dx_idvdz~e^{i[p'z-pv-\sum q_ix_i]}
\left.\frac{\delta^5S}{\delta A_{\alpha}^a(x_1)\delta
A_{\beta}^b(x_2)\delta A_{\gamma}^c(x_3)
\delta\calU(v)\delta\calU^{\dagger}(z)}\right|_0\nonumber\\
&=&ig^3\sum_{n=0}\frac{\Etilde^{(n)}(0)}{n!}\int\Pi dx_i
dydz~e^{i[p'z-\sum q_ix_i]}\delta(y-z)
\left(-\partial^2\right)^n\left\{e^{-ipy}
P[L^{a\alpha}_1L^{b\beta}_2L^{c\gamma}_3]\right\}.
\end{eqnarray}
Both $n=0$ and $n=1$ terms vanish. For $n=2$ term, we use
\begin{eqnarray}
&&\left(-\partial^2\right)^2\left\{e^{-ipy}
P[L^{a\alpha}_1L^{b\beta}_2L^{c\gamma}_3]\right\}\nonumber\\
&=&e^{-ipy}\left[(p^2)^2-4p_{\sigma}p_{\rho}\partial^{\sigma}\partial^{\rho}
+4ip^2p_{\rho}\partial^{\rho}-2p^2\partial^2\right.\nonumber\\
&&\left.-2ip_{\mu}\partial^{\mu}\partial^2-2ip_{\rho}\partial^2\partial^{\rho}
+(\partial^2)^2\right]P[L^{a\alpha}_1L^{b\beta}_2L^{c\gamma}_3],
\end{eqnarray}
and note that only terms with three or more derivatives can
contribute because of the $\delta(y-z)$ and that for the same
reason only those without $L_i$ after differentiation survive.
Extracting out
$(-i)ig^3\frac{1}{2!}\Etilde^{(2)}(0)(2\pi)^4\delta(p'-p-\sum
q_i)$, the result is
\begin{eqnarray}
&&\left[T^c\{T^a,T^b\}(q_{123}+q_{12}+2p)^{\gamma}+
\{T^a,T^b\}T^c(q_3+2p)^{\gamma}\right]g^{\alpha\beta} +2~{\rm
perms}.
\end{eqnarray}
Since the non-$g$ terms have not appeared, we have to compute
explicitly the $n=3$ term. This is the most tedious part for the
vertex as it involves derivatives up to the sixth order. One
should be very careful that derivatives do not commute because of
the path ordering operation. We skip the detail of the calculation
but recording the result. Leaving aside the common factors
$(-i)ig^3\frac{1}{3!}\Etilde^{(3)}(0)(2\pi)^4\delta(p'-p-\sum
q_i)$, the $n=3$ term is
\begin{eqnarray}
&&\left[\left(
T^c\{T^a,T^b\}[p^2+(p+q_{12})^2+(p+q_{123})^2](q_{123}+q_{12}+2p)^{\gamma}
\right.\right.\nonumber\\
&&\left.\left.+
\{T^a,T^b\}T^c[p^2+(p+q_{3})^2+(p+q_{123})^2](q_3+2p)^{\gamma}\right)
g^{\alpha\beta}+2~{\rm perms}\right]\nonumber\\
&+&\left[
T^aT^bT^c(q_{123}+q_{23}+2p)^{\alpha}(q_{23}+q_3+2p)^{\beta}(q_3+2p)^{\gamma}
+5~{\rm perms}\right].
\end{eqnarray}
Now we assume the above structure is valid for the $n$-th term and
denote the coefficients of the displayed three terms as
$b_n(p,q_1,q_2,q_3)$, $c_n(p,q_1,q_2,q_3)$, $d_n(p,q_1,q_2,q_3)$.
The initial conditions are therefore
\begin{eqnarray}
&&b_2=c_2=d_3=1,\nonumber\\
&&b_3(p,q_1,q_2,q_3)=p^2+(p+q_{12})^2+(p+q_{123})^2,\nonumber\\
&&c_3(p,q_1,q_2,q_3)=p^2+(p+q_{3})^2+(p+q_{123})^2,
\end{eqnarray}
while those not listed vanish. The coefficients in the next term
are found to be (with arguments $p,q_1,q_2,q_3$ suppressed),
\begin{eqnarray}
b_{n+1}&=&p^2b_n+
\frac{(p+q_{123})^{2n}-(p+q_{12})^{2n}}{(p+q_{123})^{2}-(p+q_{12})^{2}},
\nonumber\\
c_{n+1}&=&p^2c_n+
\frac{(p+q_{123})^{2n}-(p+q_3)^{2n}}{(p+q_{123})^2-(p+q_3)^2},
\nonumber\\
d_{n+1}&=&p^2d_n+\frac{1}{(p+q_{123})^{2}-(p+q_{23})^2}\nonumber\\
&&\times
\left[\frac{(p+q_{123})^{2n}-(p+q_3)^{2n}}{(p+q_{123})^2-(p+q_3)^2}-
\frac{(p+q_{23})^{2n}-(p+q_3)^{2n}}{(p+q_{23})^2-(p+q_3)^2}\right].
\end{eqnarray}
$b_n$ and $c_n$ have the same structure as $h_n$, while $d_n$,
when multiplied by $[(p+q_{123})^{2}-(p+q_{23})^2]$, is a
difference of two series each having the structure of $h_n$. They
are worked out to yield the final answer shown in eq
(\ref{eq_Gamma3}).


\end{document}